# Hardness prediction of age-hardening aluminum alloy based on ensemble learning

Zuo Houchen[a], Jiang Yongquan[b]*, Yang Yan[b], Liu Baoying[b] and Hu Jie[a]

a State Key Labratory of Traction Power, Southwest Jiaotong University, Chengdu, China.

b School of Computing and Artificial Intelligence, Southwest Jiaotong University, Chengdu, China.

*Corresponding author(s). E-mail(s): yqjiang@swjtu.edu.cn.

**Abstract**

With the rapid development of artificial intelligence, the combination of material database and machine learning has driven the progress of material informatics. Because aluminum alloy is widely used in many fields, so it is significant to predict the properties of aluminum alloy. In this thesis, the data of Al-Cu-Mg-X (X: Zn, Zr, etc.) alloy are used to input the composition, aging conditions (time and temperature) and predict its hardness. An ensemble learning solution based on automatic machine learning and an attention mechanism introduced into the secondary learner of deep neural network are proposed respectively. The experimental results show that selecting the correct secondary learner can further improve the prediction accuracy of the model. This manuscript introduces the attention mechanism to improve the secondary learner based on deep neural network, and obtains a fusion model with better performance. The R-Square of the best model is 0.9697 and the MAE is 3.4518HV.

# 1 Introduction

Materials are the foundation of national economy and the discovery of new materials is one of the sources driving the development of modern science and technological innovation. However, materials are a complex high-dimensional and multi-scale coupling system and it is difficult to accurately describe the relationship of material composition, structure and performance [1]. In recent years, the emergence and development of artificial intelligence has greatly changed and improved the role of computers in scientific research and engineering applications. The combination of big data and artificial intelligence is called "the fourth paradigm of science" and "the fourth Industrial Revolution" [2]. In 2011, the proposal of "Materials Genome Project" marked that the formation of informatics strategy with data-driven technology appeared in materials science [3]. Material genetic engineering is a revolutionary frontier technology in the field of materials, which will revolutionize the research and development mode of materials and accelerate the process of materials from design to engineering application. The cover article published by Nature in May 2016 also showed that machine learning could fully mine valuable information hidden behind a large amount of discarded experimental data, helping researchers to predict the composition of new materials more efficiently [4]. Rely on experimental and computational simulation to produce large amounts of data for machine learning. Especially for the calculation of large-scale and high dimension data sets, machine learning can effectively identify the characteristics of the data model, extract the implicit rule and correlation of research materials [5]. Machine learning makes material research model from the traditional "experience trial and error" model to research and development model based on data driven [6].

The combination of data mining and machine learning have been applied more and more in the construction of materials research and design platform, material analysis and prediction based on big data, which accelerates the development of materials informatics [7]. For example, in the field of predicting macroscopic properties such as hardness and mechanical properties, it is difficult for traditional material design to accurately design with performance indexes as parameters. However, the one-to-one mapping of input and output of supervised machine learning training data provides the possibility for on-demand design and prediction of material properties. Gu [8] uses the convolutional neural network model to predict the mechanical properties of two-dimensional composite materials and predict the strength and stiffness of the target material. Wang [9] constructs two artificial neural network prediction models of "component-performance" and "performance-component". Wang proposes a machine learning design system with three functions of machine learning modeling, component design and performance prediction and uses this system to design copper based alloys with different components. Zhang [10] uses physical and chemical characteristic parameters such as atomic radius, electronegativity and nuclear electron distance to construct alloy factors with material composition as weights. Zhang establishes machine learning models of tensile strength and conductivity of copper alloys. Chaudry [11] uses hardness data samples of 1591 Al-Cu-Mg-x(X: Zn, Zr, etc.) alloys and material parameters such as composition, aging temperature and aging time to establish a gradient boosting regression tree model to predict the hardness of aluminum alloy under various aging conditions and R-square of the model is 0.95. Russlan [12] also builds a model to predict the hardness of aluminum alloy based on 1591 samples. Russlan combines multi-objective evolutionary algorithm and gradient boosting regression tree and the relative error of model is 3.5%.

## 2 Materials and Methods

### 2.1 Data Set

Age-hardening is the phenomenon that supersaturated solid solution increases the strength and hardness of alloy in the aging process, which is of great significance in the practical application of industry, and many alloys are designed accordingly. In this manuscript, Al-Cu-Mg-X (X: Zn, Zr, etc.) alloy is selected as the base material, and the selected characteristic data include the composition, aging conditions (time and temperature) and related properties (hardness) of the alloy. This data set is derived from the work carried out by literature [11]. These data are obtained from papers published in top journals in the field of metallurgy, such as Acta Materialia, Scripta Materialia, Journal of Alloys and Compounds and Materials Science & Engineering A. These data are of high quality, authentic and can be used as data sets for machine learning. The composition, aging time and aging temperature of the aluminum alloy are taken as the input characteristics. Hardness of the aluminum alloy is taken as the output characteristics. The data set contains 1591 records, which covers 71 materials.

## 2.2 Methods

After years of development, many types of algorithms have emerged in machine learning. How to select an effective model is an important problem that machine learning has to face in practical application, because it is impossible to have a model that can solve all problems. Most models are proposed for a certain kind of problem or even a certain problem. To solve this problem, we use the automatic machine learning framework auto_ml to make preliminary predictions and screen out the required models. The purpose of automatic machine learning [13] is to automatically discover and construct relevant features, automatically design the model structure, set its optimal parameters and optimization algorithm. Finally, automatic machine learning can provide an end-to-end solution. In this paper, an ensemble learning method is proposed which combines automatic machine learning auto_ml and model fusion strategies, and this method is used to predict the hardness of aluminum alloy after aging treatment. The process of this experiment is shown in Figure 1. The data set used is the data used in literature [11]. The primary learner is screened by auto_ml and the hyper parameters of models are optimized by random search and grid search. We use averaging method and learning method to fuse primary learners. The average methods use absolute average strategies and weighted average strategies. The secondary learners of the learning method use auto_ml to filter and introduce attention mechanism to optimize model. Finally, by comparing the prediction accuracy of different models, the model with the best prediction effect is found.

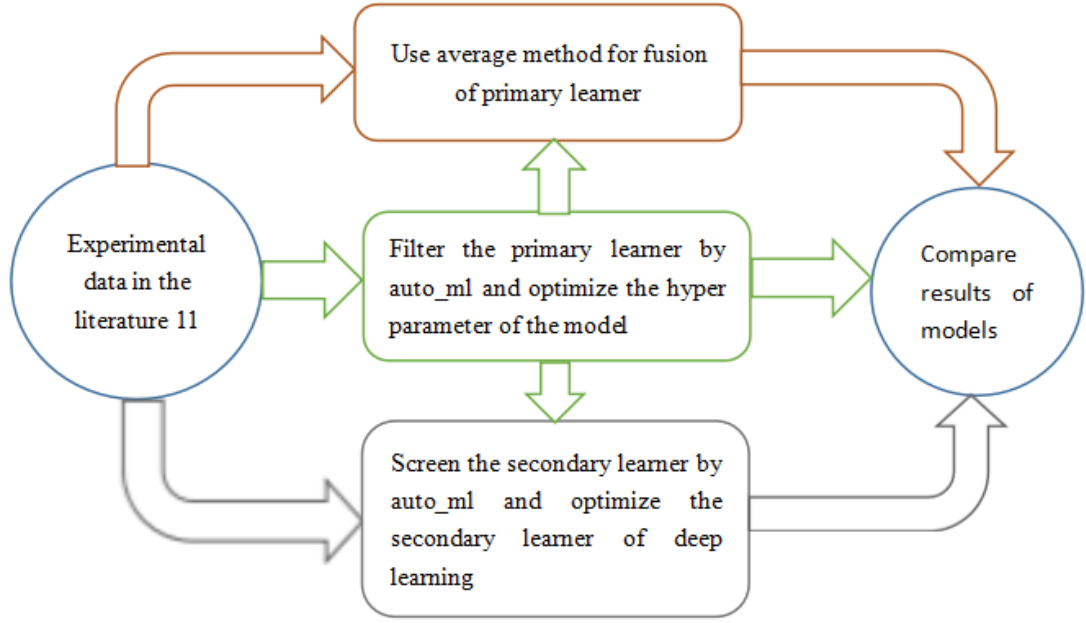

Figure 1. Experimental workflow

## 3 Results and Discussion

### 3.1 Model evaluation criteria

R-Square and MAE are used to evaluate the prediction results of the model. The value range of R-Square is generally between 0 and 1, and the closer it is to 1, the better the model effect is. However, R-Square will be negative when the fitting effect is too poor. Its definition is shown in Formula (1). In addition, this experiment also uses MAE (Mean Absolute Error) to evaluate the experimental model. MAE is the average of the absolute error between the predicted value and the real value, and the closer it is to 0, the better the model effect is. Its definition is shown in Formula (2). In equations (1) and (2), n is the number of samples, $\omega_i$ is the weight coefficient, $\bar{y_1}$ is the average value of real data, $y_i$ $f(x_i)$ is real data, and is predicted data.

$$R\text{-}Square = 1 - (\sum_{i=1}^{n}\omega_i(f(x_i)-\bar{y_1})^2 / (\sum_{i=1}^{n}\omega_i(y_i-\bar{y_1})^2 \tag{1}$$

$$MAE = (1/n)\sum_{i=1}^{n}|y_i - f(x_i)| \tag{2}$$

## 3.2 Screening and Hyper Parameter Optimization of Preliminary Models

Auto_ml is used in the experiment to conduct preliminary screening of the model and the suitable machine learning models are selected. The preliminary screening results of the model are shown in Table 1 and the evaluation criterion is R-Square. We bolder the prediction effect of the better models in the Table 1. According to preliminary screening based on auto_ml, extreme random tree model (remember to Extra Trees), gradient boosting regression model (remember to GBR), random forest model (remember to Random Forest), gradient boosting decision tree model based on LightGBM algorithm (remember to LightGBM) and XGBoost algorithm (remember to XGBoost) have better prediction effect.

Table 1 Preliminary screening results based on auto_ml

| Model | R-Square |
| --- | --- |
| ARD | 0.5890 |
| Ada Boost | 0.7801 |
| Bayesian Ridge | 0.5681 |
| Extra Trees | **0.9524** |
| GBR | **0.9574** |
| Lasso | 0.4282 |
| Linear | 0.5796 |
| Logistic | 0.0943 |
| Passive Aggressive | 0.4408 |
| Random Forest | 0.9467 |
| Ridge[57] | 0.4760 |
| Deep Learning | 0.6506 |
| LightGBM | **0.9523** |
| XGBoost | **0.9574** |

The selected models need to set hyper parameters to improve the performance of models. To solve this problem, two hyper parameter optimization strategies are selected, including of random search and grid search. Random search is mainly used for the search of huge hyper parameter range. Grid search is to adjust the hyper parameters by step in the specified hyper parameter range and uses the adjusted hyper parameters to train models to find the hyper parameters with the highest accuracy on the verification set from all the hyper parameters. However, these search strategies have their own limitations. Random search will not go through all possible parameter collection, so the final result may only be a relatively good solution. Grid search may consume too much time when the search scope is too large and even more serious may cause the program crash. Aiming at the selection of hyper parameters, this part fuses two kinds of hyper parameter optimization strategies. We first use a large range of random strategy to find a relatively optimal point and then use a small range of grid search. We mainly target max_depth and iteration times (n_estimators), because the selection of these two hyper parameters has a great influence on the prediction results of the model. In the experiment, we set the maximum depth of the tree as 30, step size as 3, and the maximum number of iterations as 1000, step size as 10. Firstly, random search is used to select 300 times of hyper parameters, and then grid search is used to perform exhaustive search near the best obtained hyper parameter until the best hyper parameters is found. The search range we set in the experiment is the size of the random search step. Assuming that the depth value finally obtained is 15 and the number of iterations is 500, the hyper parameters of the optimal model will be generated in the depth from 12 to 18 and the number of iterations from 490 to 510. The hyper parameters and prediction results of the selected models are shown in Table 2. We bold the prediction effect of the best models in the Table 2. From Table 2, we can see that LightGBM model and Extra Trees model achieve the best results on R-Square and MAE respectively.

Table 2 Hyper parameters and prediction results of models

| Model | Hyper parameters | R-Square | MAE(HV) |
| --- | --- | --- | --- |

|  | max_depth | n_estimators |  |  |
|---|---|---|---|---|
| Extra Trees | 19 | 93 | 0.9653 | **3.5462** |
| GBR | 5 | 483 | 0.9665 | 3.7615 |
| RandomForest | 27 | 114 | 0.9605 | 3.9730 |
| XGBoost | 17 | 806 | 0.9617 | 3.8855 |
| LightGBM | 6 | 860 | **0.9679** | 3.7845 |

## 3.3 Ensemble Models Based on Average Fusion Strategies

We check the predicted values of these five models and select several predicted data for presentation. The results are shown in Table 3. The best predicted hardness values are bolded and the worst predicted hardness values are underlined in Table 3. It can be seen from Table 3 that no matter which model is used, there are values with very accurate prediction and values with large prediction deviation.

Table 3 Hardness prediction results of models

| Hardness | Extra Trees | GBR | Random Forest | XGBoost | LightGBM |
|---|---|---|---|---|---|
| 94 | 93.0850 | **94.4726** | 92.8860 | 93.0363 | 96.2640 |
| 97 | 97.8586 | 94.3443 | 95.3772 | **97.1013** | 96.8583 |
| 104 | 105.6229 | 106.1266 | 106.3794 | 105.8741 | **104.2025** |
| 115 | 114.8841 | 116.7124 | **114.9810** | 113.0902 | 113.5329 |
| 118 | 115.9409 | **117.8152** | 117.4737 | 117.0676 | 116.9626 |
| 128 | 129.3938 | 129.1912 | 130.0351 | **128.8134** | 130.3741 |
| 130 | **130.6989** | 132.1759 | 136.8246 | 133.3669 | 135.7790 |
| 169 | **168.8602** | 169.2453 | 169.4123 | 169.4018 | 170.2219 |
| 172 | 169.5806 | 174.0944 | 169.6053 | 170.9956 | **171.7291** |
| 181 | 186.5591 | 182.8237 | **181.4737** | 185.9953 | 189.2366 |

In view of this situation, the experiment chooses corresponding strategies to integrate these five models into a new model, so as to obtain higher prediction accuracy. In the ensemble regression models, the common methods are absolute average strategy and weighted average strategy. Absolute average strategy is easy to understand. In the experiment, models optimized by hyper parameters are used as the primary learners, and the prediction results of these five models are added and divided by the number of learners. However, the fusion strategy of absolute average strategy is too violent and depends very much on the prediction accuracy of primary learners. Once primary learners have poor prediction effect, the fusion model using absolute average strategy will not have a good prediction result, even prediction accuracy is lower than the best prediction model of primary learners. We also use weighted average strategies in the experiment. Each model is multiplied by the corresponding weight and added together, and the final result is the output result of the fusion model. The first weight used in the experiment is based on the R-Square of the model on the training set. The weight of each model is the proportion of the corresponding R-square to the sum of the r-square of the five models. The second weight used in the experiment is based on the MAE of the model on the training set, but MAE is different from R-Square. The higher R-Square is, the better the model effect is, but the higher MAE is, the worse the model effect is. Therefore, the MAE weight used in the experiment is the ratio of the MAE average of the other four models to MAE sum of the five models. Using this algorithm, the lower the MAE of the model is, the higher the weight ratio is. The results of three averaging strategies are shown in Figure 2. It can be seen from the Figure 2 that the prediction accuracy of three fusion models is higher than single model, and the fusion model based on MAE weighted average has the best prediction effect.

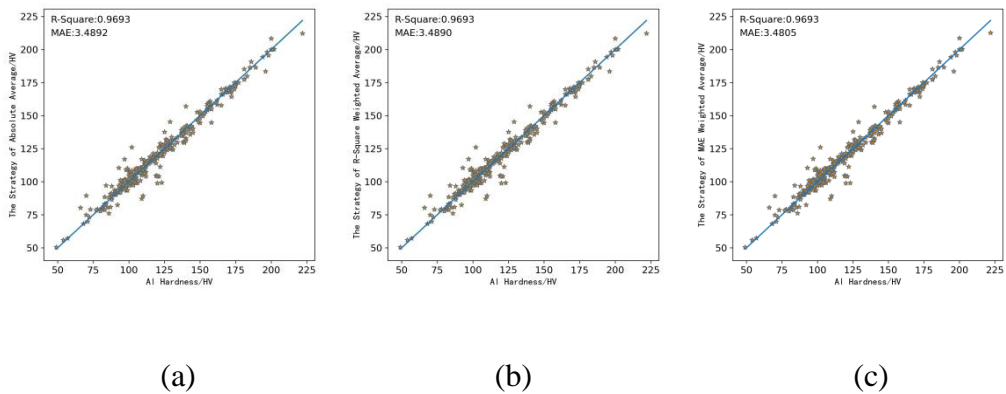

(a)                                    (b)                                    (c)

Figure 2. (a) is the result of absolute average strategy, (b) is the result of weighted average strategy based on R-Square, (c) is the result of weighted average strategy based on MAE

## 3.4 Ensemble Models Based on Learner Fusion Strategies

In the work, most of the fusion model schemes are realized based on the average method, and the model fusion strategy based on the learning method is rarely used, even several classical ensemble learning models adopt the absolute average strategy or weighted average strategies for model integration. On the one hand, the fusion model based on average strategy can improve the prediction accuracy in most cases and achieve expected effect. On the other hand, faced with a wide variety of machine learning models, users do not know which model to choose as the secondary learner. Blindly trying models will only lead to low efficiency. On how to use the model of integration of learning strategies, this part gives a solution based on automatic machine learning framework auto_ml. The process that we give is to use auto_ml model to screen primary learners and secondary learners. Firstly, we use auto_ml and hyper parametric search strategies to optimize primary learners. Then we use average strategies and learning method strategy to fuse primary learners. It is worth noting that secondary learners of the learning method strategy are filtered by auto_ml. By introducing automatic machine learning, the process of model fusion becomes very simple and the screening results are shown in Table 4. It can be seen from Table 4 that the prediction accuracy of the fusion model based on ARD, Bayesian Ridge, Extra Trees, Linear, Random Forest and Ridge learner is higher than primary learners. The secondary learner of the best model based on the learning method is the Lasso learner and its prediction result is shown in Figure 3 (a). Through this solution based on auto_ml, model fusion can save a lot of time. The prediction accuracy of fusion model based on the learner selected by this scheme is better than model based on MAE weighted average.

We analyze the relevant source code of auto_ml and find that its deep learning network structure is fixed and auto_ml focuses on the optimization of input characteristics. This may lead to the poor performance of deep learning provided by automatic machine learning in this problem. Therefore, we rebuild a 9-layers deep fully connected network learner in the work and the prediction effect is shown in Figure 3 (b). Experimental result shows that the performance of the fusion model based on deep network learner is better than model based on Lasso learner.

The neural network provided by automatic machine learning auto_ml will process the input features, increasing the weight of key features and reducing the weight of non-key features, which is similar to the idea of attention mechanism. Therefore, in order to further improve the prediction accuracy of the model, the experiment introduces an attention mechanism to optimize the deep network in the secondary learner to improve the weight of key information and reduce the interference of irrelevant information. In order to find the optimal model structure, the experiment adds 1~4 attention layers into the model respectively and the prediction results are shown in Table 4. The predictions of the best models are bolded, and those of the worst are underlined in Table 4. According to experimental results, the fusion model achieves the best prediction performance with MAE of 3.4518HV and R-Square of 0.9697, which exceeds the prediction accuracy of all previous models. The experiment also reproduces the models in literature [11] and [12] and uses the same test set for testing. The results are in Table 4 and we can the prediction accuracy of the optimal model proposed by the experiment exceeds models in literature [11] and [12]. The result of best attention network is shown in Figure 3 (c). The structure of best attention network is shown in Figure 3 (d). The process is that five primary learners make predictions first and then their prediction results are used as the input features of the secondary learners to train the model. The number of neurons in each neural layer is recorded in parentheses. The symbol $\otimes$ after input layer and hidden layer 2 represents the dot product of the features of this layer and the features of the attention layer. In this model, primary learners are extreme random tree, random forest, gradient boosting regression model and two kinds of gradient boosting regression tree models, and the secondary learner is attention network.

Table 4 The results of screened secondary learners and literature [11] [12]

| Model | MAE | R-Square |
| --- | --- | --- |
| Model in Literature [11] | 4.5212 | 0.9503 |
| Model in Literature [12] | 4.1814 | 0.9604 |
| ARD | 3.5153 | 0.9670 |
| Ada Boost | 4.1671 | 0.9619 |
| Bayesian Ridge | 3.5175 | 0.9670 |
| Extra Trees | 3.4947 | 0.9658 |
| GBR | 3.5896 | 0.9647 |
| Lasso | 3.4652 | 0.9675 |
| Linear | 3.5199 | 0.9669 |
| Logistic | 10.3542 | 0.8340 |
| Passive Aggressive | 3.5602 | 0.9665 |
| Random Forest | 3.5059 | 0.9648 |
| Ridge | 3.4806 | 0.9692 |
| Deep Learning | <u>14.1663</u> | <u>0.4897</u> |
| LightGBM | 3.8036 | 0.9621 |
| XGBoost | 3.6379 | 0.9628 |
| Optimized Deep Learning | 3.4612 | 0.9693 |
| Network based on one attention layer | 3.6099 | 0.9640 |
| Network based on two attention layers | **3.4518** | **0.9697** |
| Network based on three attention layers | 3.6348 | 0.9678 |
| Network based on four attention layers | 3.7558 | 0.9613 |

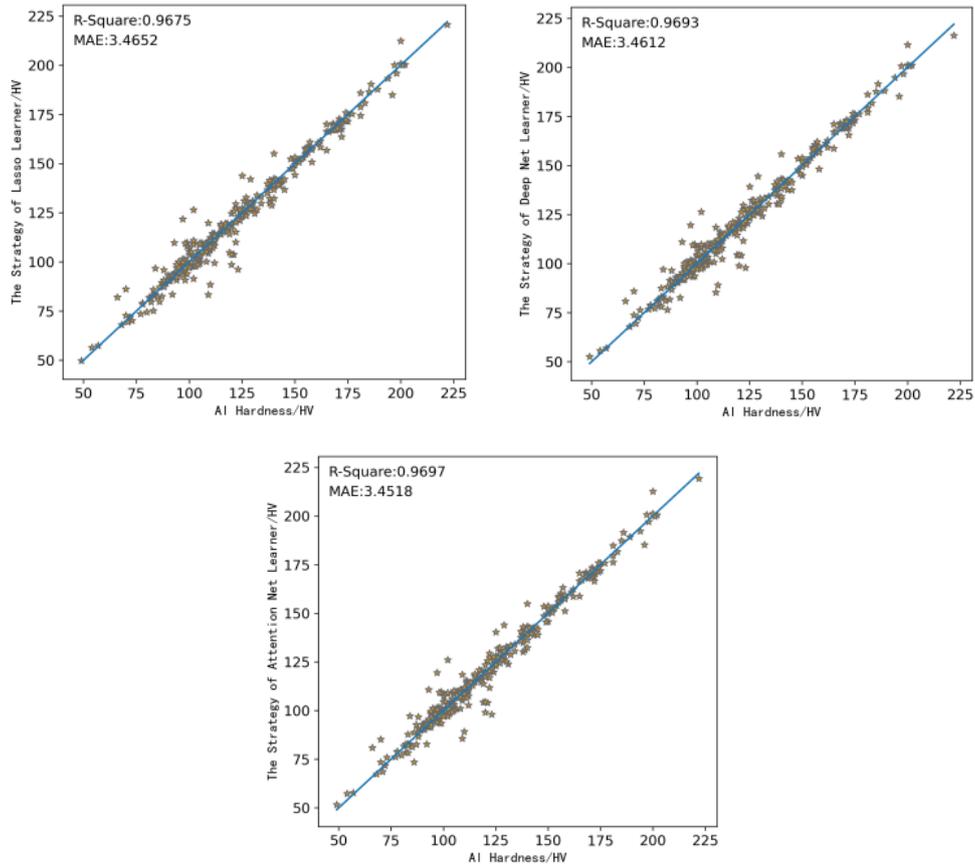

(a)                                          (b)                                          (c)

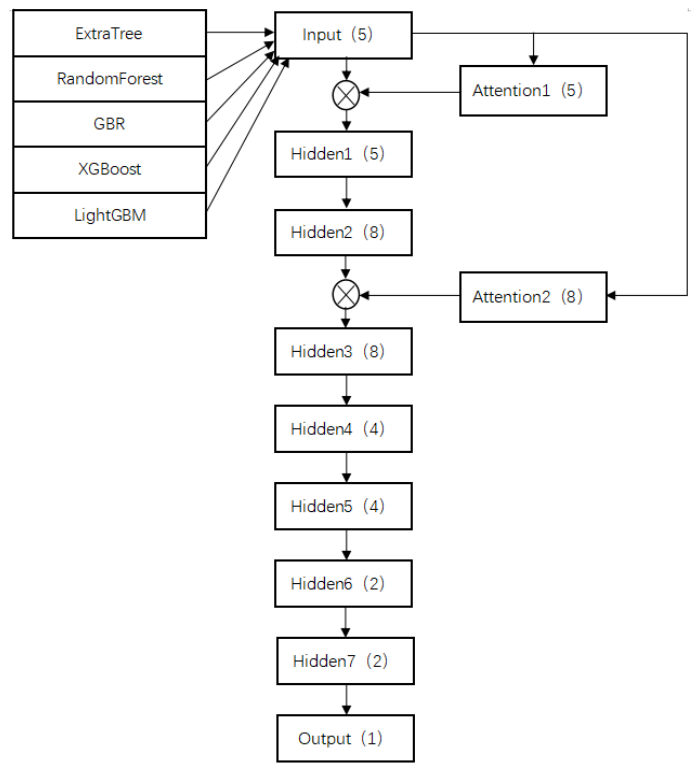

(d)

Figure 3. (a) is the best ensemble model based on auto_ml, (b) is the ensemble model

based on deep network, (c) is the best ensemble model based on attention net, (d) is the structure of ensemble model with attention mechanism

## 3.5 Discussion

We analyze the prediction results of the best model and extract the materials with good prediction and those with bad prediction for analysis. The extraction results are shown in Table 5 and the numbers after the elements indicate the proportion of elements in the material. It can be seen from Table 5 that the accurately predicted values are almost the same as the experimental values, and the error of the poorly predicted values is relatively large. When analyzing the predicted value with large error, it is found that Al92.3-Cu0.45-Mg5.2-Zn2 appears for many times in Table 5, so this material is not suitable for the ensemble learning model in this part. The predicted data of Al97.2-Cu1.1-Mg1.7 corresponds to the turning point of hardness change in the age-hardening process and Al90.6-Cu3.8-Mg0.26-Si5-Fe0.2-Ti0.08 corresponds to the stable point of hardness in the age-hardening process. After aging treatment, the hardness of the material will first increase and then decrease, and eventually remain unchanged. The reason of predicting these two materials poorly is that these two data corresponds to the special time point in the process of age-hardening, but test data do not have enough information to support.

Table 5 The hardness prediction error of the best model

| Material | Temperature/K | Time/min | Error/HV |
| --- | --- | --- | --- |
| Al93.75-Cu0.45-Mg5.2-Zn0.6 | 453 | 50 | 0.0052 |
| Al95.7-Cu2.5-Mg1.5-Si0.25 | 473 | 1100 | 0.0140 |
| Al93.5-Cu4.8-Mg0.72-Si0.04-Zr0.16 -Mn0.27-Ag0.36-Fe0.02-Ti0.04 | 438 | 3600 | 0.0189 |
| Al90.3-Cu0.15-Mg5.1-Si0.15-Zn3 -Zr0.15-Mn0.8-Fe0.2-Ti0.07-Cr0.03 | 363 | 4200 | 0.0238 |
| Al96.3-Cu1.89-Mg1.56-Mn0.21 | 298 | 240 | 0.0392 |
| **Al92.3-Cu0.45-Mg5.2-Zn2** | 453 | 400 | 24.9195 |
| Al97.2-Cu1.1-Mg1.7 | 423 | 1666 | 24.0979 |
| Al90.6-Cu3.8-Mg0.26-Si5-Fe0.2-Ti0.08 | 448 | 2940 | 23.4307 |
| **Al92.3-Cu0.45-Mg5.2-Zn2** | 453 | 600 | 22.4887 |
| **Al92.3-Cu0.45-Mg5.2-Zn2** | 453 | 8000 | 20.9210 |

## 4 Conclusion

For the hardness of aluminum alloy after age-hardening, this paper adopts the model fusion strategy to conduct experiments. Based on the automatic machine learning framework auto_ml, a fusion model solution is proposed. The process is to use auto_ml to screen models and optimize the hyper parameters of models. Then the model is fused using averaging strategies and learning method strategy based on auto_ml. In this manuscript, attention mechanism is introduced to improve the secondary learner based on deep neural network and a fusion model with better performance is obtained. The R-Square of the model is 0.9697 and MAE is 3.4518hV, which exceeds the accuracy of models used in literature [11] and [12]. Experimental result confirms the feasibility and effectiveness of our method. This model can be used to screen unknown materials and provide direction for experiments. For example, if you want to find aluminum alloy materials with higher hardness, you can create a search space by simulating the composition and aging conditions of materials. The results of the search space may provide direction for research and avoid blind experiments to some extent.

## CRediT authorship contribution statement

**Zuo Houchen:** Software,Visualization,Writing. **Jiang Yongquan:** Conceptualiza-tion,Methodology,Supervision. **Yang Yan:** Funding acquisition,Project administrati-on. **Liu Baoying:** Revise. **Hu Jie:** Suggestion.

## Declaration of Competing interest

The authors declare that they have no known competing financial interests or personal relationships that could have appeared to influence the work reported in this paper.

## Acknowledgements

The research work was supported by the National Natural Science Foundation of China (No.61976247) and the Fundamental Research Funds for the Central Universities (No.2682021ZTPY110).

17    Hardness prediction of age-hardening aluminum alloy based on ensemble learning